\documentclass{article}

\usepackage{hyperref}       
\usepackage{url}            
\usepackage{booktabs}       
\usepackage{amsfonts}       
\usepackage{nicefrac}       
\usepackage{microtype}      
\usepackage{braket}         

\usepackage{graphicx}
\usepackage{amsmath}
\usepackage{amsthm}
\usepackage{amssymb}
\usepackage{mathtools}
\usepackage{dsfont}
\usepackage[backend=biber, bibencoding=utf8, sorting=none]{biblatex}
\usepackage{inputenc}
\usepackage{hyperref}
\usepackage{clrscode3e}
\usepackage{caption}
\usepackage{multicol}
\usepackage{listings}

\usepackage{tikz}
\usetikzlibrary{bayesnet}

\usepackage{fullpage}
\usepackage{float}
\usepackage[affil-it]{authblk}

\usepackage[colorinlistoftodos]{todonotes}

\DeclareMathOperator{\defeq}{\vcentcolon=}

\DeclareUnicodeCharacter{21E4}{-}

\newcommand{\lang}[1]{\ensuremath{\mathsf{#1}}}

\newtheorem{definition}{Definition}
\newtheorem{claim}{Claim}

\hypersetup{
colorlinks=true,
linkcolor=blue,
filecolor=black,
citecolor=blue,
urlcolor=black
}

\title{A Review of zk-SNARKs}
\author{Thomas Chen\footnote{Columbia University, thomas.chen@columbia.edu}, Hui Lu\footnote{Columbia University, abby.lu@columbia.edu}, Teeramet Kunpittaya\footnote{Columbia University, teeramet.kunpittaya@columbia.edu}, and Alan Luo\footnote{Columbia University, al3856@columbia.edu}}

\date{}

\begin{document}
\maketitle

\begin{abstract}
A zk-SNARK is a protocol that lets one party, the prover, prove to another party, the verifier, that a statement about some privately-held information is true without revealing the information itself. This paper describes technical foundations, current applications, and some novel applications of zk-SNARKs. Regarding technical foundations, we go over the Quadratic Arithmetic Program reduction and the Pinocchio protocol. We then go over financial security applications like Zcash and Tornado Cash, and zk-Rollup applications like zkEVM and Darkforest. We propose novel zk-SNARK protocols for private auctions and decentralized card games on the blockchain, providing code for the proposed applications. We conclude by touching on promising zk-SNARK innovations, such as zk-STARKs.

\end{abstract}

\section{Introduction}

zk-SNARKs (Zero-Knowledge Succinct Non-interactive Arguments of Knowledge) have gained increasing amounts of interest due to their uses in scaling blockchain throughput and providing secure ways to perform transactions \cite{zcashpaper, tornado, zkrolluparchitecture, darkforest}. The goals of this review are to provide a technical understanding of zk-SNARKs, to describe their key current applications, and to propose zk-SNARK protocols for two novel applications. 

First, we provide a history of the development of zk-SNARKs in section (\ref{section:history}). In section (\ref{section:technical}), we give a detailed dive into the technical workings of zk-SNARKs -- in particular the Pinocchio protocol \cite{PHGR}. Sections (\ref{section:financeApps}) and (\ref{section:rollupApps}) are dedicated to financial and rollup applications of zk-SNARKs respectively, two of the most important current applications. In section (\ref{section:newApps}), we propose novel (to our knowledge) circuits that use zk-SNARKs in two applications: private auctions and decentralized card games. We also provide code for these proposed applications in the appendix (\ref{appendix}). Finally, in (\ref{section:future}) we conclude the review by describing new innovations on zk-SNARKs: zk-STARKs and recursive SNARKs.

As a motivating example, zk-SNARKs are useful for blockchain rollups where the prover wants to keep the witness of their computation a secret. Suppose Alice runs some computation off-chain, and in her computation, she obtains some secret witness, such as the details of a ZCash transaction, or the location of a newly-generated planet in the zk-game, Dark Forest \cite{darkforest}. Alice would like the blockchain to commit the state changes her computation made, without revealing her secret witness and without the chain rerunning her entire computation on-chain. To this end, she would like a protocol where she, the prover, can convince the chain, the verifier, that a statement is true, but reveal nothing else. Such proof systems are called zero knowledge.

The following is a concrete example of a zero knowledge proof system, for the graph $3$-coloring problem (\lang{G3COL}). A prover $P$ and verifier $V$ both know a graph $G$. $P$ wants to convince $V$ that she has a valid $3$-coloring of $G$, without disclosing her coloring. The following protocol can be used \cite{graph3color}:

\begin{enumerate}
    \item $P$ randomly permutes the colors of the vertices of $G$ (e.g. all red vertices becomes blue, all blue become green, and all green become red). Then, $P$ commits the colors of all vertices using a commitment-scheme.
    \item $V$ randomly chooses an edge $e_{i, j} \in E$ and asks $P$ for the colors of the two vertices, $i$ and $j$.
    \item $P$ reveals the committed colors of $i$ and $j$ to $V$.
    \item $V$ repeats steps (1) -- (3) $n$ times and accepts if and only if each of his $n$ queries were answered with two distinct colors.
\end{enumerate}

If $P$ has a true $3$-coloring of $G$, she will convince $V$ with probability $1$. In contrast, if $P$'s coloring contains at least one edge whose vertices are the same color, she will convince $V$ with probability at most $p = (1 - 1/E)^n$. Thus, after a sufficient number of repetitions, $n = O(E \ln \frac{1}{\delta})$, $V$ is confident that $P$ has a true graph coloring with probability $1 - \delta$.

The above is an interactive proof system, where two parties' interactions enable them to recognize some language, $L$. The proof system should be such that any statement $x \in L$ has a proof causing $V$ to accept it, while any statement $x \notin L$ has no proof that causes $V$ to accept it. The protocol should also not reveal $x$ to $V$–– only convince $V$ that $P$ has a member of $L$. These three key properties of zero-knowledge proof systems are defined below. Formal definitions can be found in the appendix (\ref{appendix}).

\begin{enumerate}
    \item Completeness: Every statement with a valid witness has a proof that convinces the verifier.
    \item Soundness: A statement with an invalid witness does not have a proof that convinces the verifier.
    \item Zero-knowledge: The proof only conveys information about the validity of the statement and nothing about the prover's witness.
\end{enumerate}

\subsection{History}\label{section:history}

Zero-Knowledge interactive proof protocols were the starting point of zk-SNARKs. Goldwasser, Micali, and Racknoff \cite{Goldwasser1989} first introduced sound, complete, and zero-knowledge interactive proof systems for \lang{Quadratic-Residuosity} and \lang{Quadratic-Nonresiduosity}, two languages in not known to be efficiently recognizable \cite{Goldwasser1989}.

Subsequently, Blum, Feldman, and Micali developed non-interactive zero-knowledge (NIZK) proofs \cite{Blum1991}, using shared randomness to create a common reference string (CRS). Unlike the protocol for \lang{G3COL}, a non-interactive protocol requires just one message to be sent from prover to verifier, cutting out the back-and-forth interaction. This feature is especially important in blockchain rollups, so that the prover does not need to be around every time someone wants to verify their proof. 

The next key innovation was in reducing the size of the proof sent. Kilian \cite{Kilian} gave an interactive protocol for sublinear zero-knowledge arguments that sent fewer bits than the size of the statement to be proved. Micali then created the first sublinear-size NIZK proofs \cite{Micali2000}.

The next goal was to make proofs constant size. Groth, Ostrovsky, and Sahai \cite{GOS06, Groth2011} introduced pairing-based NIZK proofs. Using these ideas, Groth developed the first constant-size NIZK arguments system for \lang{Circuit-SAT}, an \lang{NP}-complete language \cite{Groth-10}. Groth's protocol used pairings to efficiently check polynomial relationships between the discrete logs (the decoded versions) of the prover's encoded messages, without needing to know the decoded messages themselves. The proof is created so that checking a particular polynomial relationship between elements of the proof is equivalent to verifying that prover has a satisfying assignment for the circuit, a core idea in current zk-SNARKs.

Continuing on with the idea of verifying polynomial equations, Gennaro, Gentry, Parno, and Raykova \cite{GGPR13} introduced the now-popular Quadratic Span program (QSP) and Quadratic Arithmetic Program (QAP) reductions, where an \lang{NP}-statement is reduced to a statement about QSPs. Their method's proof size, proof generation complexity, and CRS setup complexity were much more efficient than Groth's constant-size NIZK arguments system \cite{GGPR13}.

The Pinocchio protocol used QAP reductions to create a constant-size NIZK arguments protocol with CRS and proof generation time linear in the circuit size. Moreover, verification in Pinocchio is takes less than $10$ ms, a $5-7$ order of magnitude improvement from previous work \cite{PHGR}. Pinocchio is protocol Zcash first used \cite{zcash, PHGR}. Subsequently, the Groth-$16$ protocol \cite{Groth2016} improved on Pinocchio's proof size and verifier complexity, having just $3$ group elements in its proof. It has since become the state-of-the-art, used in Zcash and Circom \cite{circom}.

The crucial improvements to make zero-knowledge proofs non-interactive and succinct made them practical for blockchains. In contrast, the example graph $3$-coloring proof system is interactive, and the proof size is linear in $E$. We will now describe how a zk-SNARK achieves all these properties, for proving \lang{NP}-statements (statements in \lang{NP}, e.g. that a graph is in \lang{G3COL}) about a wide range of computations.

\section{zk-SNARK Construction}\label{section:technical}

\subsection{Preliminaries}

\begin{figure}[H]
\centering
\includegraphics[width=10cm]{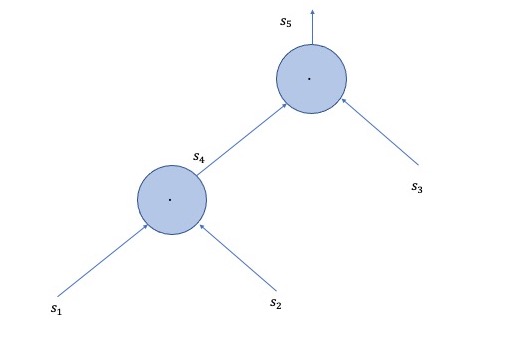}
\caption{Arithmetic circuit for $f(s_1, s_2, s_3) = (s_1 \cdot s_2) \cdot s_3$}
\label{figure:arith-circuit-clean}
\end{figure}

\begin{definition}(Arithmetic Circuit) A directed acyclic graph, with nodes as addition and multiplication gates, and edges as wires, over a finite field $\mathbb{F}_p$. Wires connect the outputs of one gate to the inputs of another. Each gate has two inputs and one output wire, and the circuit has one final output wire. Figure (\ref{figure:arith-circuit-clean}) is an example of an arithmetic circuit.
\end{definition}

For an $m$-gate, $n$-wire circuit, define a witness $s = (s_1, s_2, ..., s_n)$ to be the assignments to the $n$ wires of the circuit such that each gate's inputs and outputs satisfy the constraint defined by the gate operation. An $m$-gate, $n$-wire arithmetic circuit defines a relation over witnesses $s = (s_1, s_2, ..., s_n)$ such that for some constants $\{ u_{i,q}, v_{i,q}, w_{i,q}\}_{1 \leq i \leq n, 1 \leq q \leq m}$,
\begin{align}
\sum_{i = 1}^{n} s_i u_{i, q} \cdot \sum_{i = 1}^{n} s_i v_{i, q} = \sum_{i = 1}^{n} s_i w_{i, q} \quad \forall 1 \leq q \leq m
\end{align}

The constraints above are a set of $m$ rank-$1$ constraints, which model the relationships a circuit's multiplication gates enforces over its input and output wires. An example of a particular rank-$1$ constraint is $s_1 \cdot s_2 - s_3 = 0$, corresponding to a multiplication gate that takes as input $s_1, s_2$ and outputs $s_3$. A set of $m$ rank-$1$ constraints can be generalized into a quadratic arithmetic program, which makes QAPs natural to reduce arithmetic circuits to.

\begin{definition}(Quadratic Arithmetic Program) Pick target points $r_1, r_2, ... r_m \in \mathbb{F}_p$. Define $t(x) = \prod_{q = 1}^{m} (x - r_q)$. Further, let $u_i(x), v_i(x), w_i(x)$ be degree $m - 1$ polynomials such that for $1 \leq i \leq n, 1 \leq q \leq m$
\begin{align}
u_i(r_q) &= u_{i,q}\\
v_i(r_q) &= v_{i,q}\\
w_i(r_q) &= w_{i,q}
\end{align}

Then, a Quadratic Arithmetic Program is a relation over $s = (s_1, s_2, ..., s_n)$ such that
\begin{align}
\sum_{i = 1}^{n} s_i u_{i}(x) \cdot \sum_{i = 1}^{n} s_i v_{i}(x) - \sum_{i = 1}^{n} s_i w_{i}(x) \equiv 0      \quad    (mod \quad t(x))
\label{eq:qap}
\end{align}
\end{definition}

Each target point corresponds to a gate in the circuit. For each target point $r_q$, the QAP constructs $3n$ polynomials that when evaluated at $r_q$, yield the $3n$ constants of the $q$th gate's rank-$1$-constraint. The QAP expresses the $m$ rank-$1$ constraints as a QAP as a single equation over the polynomials. Thus, checking polynomial equality in Equation (\ref{eq:qap}) is equivalent to checking the m rank-1 constraints simultaneously. We will eventually see that this is crucial to make our proofs succinct.

Checking particular equality and divisibility conditions about QAPs is key to our zk-SNARK protocols. To do these checks in an encrypted manner, we introduce the following two concepts:

\begin{definition}(Homomorphic Encoding) An injective homomorphism $E: \mathbb{F}_p \rightarrow G$ such that it is hard to find $x$ given $E(x)$.
\end{definition}

For a cyclic group $G$ of prime order $p$ and multiplication as the group operation, we will use the encoding $E(a) = g^a$, for a generator $g$ of $G$. This is homomorphic and injective. Moreover, by the hardness of the discrete logarithm problem, it is also hard to find $x$ given $E(x)$.

\begin{definition}(Pairing Function) Suppose $G_1, G_2,$ and $G_T$ are groups of prime order $p$. A pairing function, or bilinear map, is a function 
\begin{align}
e : G_1 \times G_2 \rightarrow G_T
\end{align}

such that if $g, h$, are generators of $G_1$ and $G_2$ respectively, then $e(g,h) \neq 1$ is a generator of $G_T$ and $e(g^a, h^b) = e(g,h) ^{ab}$. 

\end{definition}
Below, we list some basic properties about bilinear maps.

\begin{align}
e(u, vw) &= e(g^a, h^b h^c) = e(g,h)^{a(b + c)} = e(g^a,h^b) e(g^a, h^b) = e(u,v) e(u, w)\\
e(vw, u) &= e(v,u) e(w, u)\\
e(u^x, v) &= e(g,h)^{xab} = e(u, v^x)
\end{align}

For the Pinocchio implementation, we use symmetric bilinear maps, where $G := G_1 = G_2$ \cite{PHGR}.

In addition to these preliminaries for the Pinocchio protocol, we've included more formal definitions for soundness, completeness, and zero-knowledge in the appendix (\ref{appendix}), as well as the subtle difference between a proof system and an argument system.

\subsection{Reducing Arithmetic Circuits to QAP}

Arithmetic-Circuit Satisfiability is an \lang{NP}-complete language \cite{Groth2016}. Therefore, for any \lang{NP}-computation that a prover and a verifier both agree on beforehand, they can construct an arithmetical circuit for that computation such that a witness satisfying the circuit is a witness for the original computation. It is possible to build circuits for a variety of useful \lang{NP}-computations, such as the language of valid terrains according to some terrain-generation algorithm, or the language of valid transactions according to some transaction protocol.

Thus, for an arbitrary arithmetic circuit, the prover needs a way to prove to the verifier that they have a witness, without revealing it. Since Pinocchio take as input an instance of QAP, we first must show how to reduce an arithmetic circuit to a QAP, such that an \lang{NP}-statement about satisfiability of an arithmetic circuit can be made into an equivalent \lang{NP}-statement about satisfiability of a QAP. This reduction should further be such that the witness of the arithmetic circuit is a witness of the constructed QAP. Then, verifying that the prover has a witness for the QAP is equivalent to verifying the original computation was done correctly. 

\subsubsection{An Example Reduction}

Consider the arithmetic circuit in figure (\ref{figure:arith-circuit-clean}). Both verifier $V$ and prover $P$ know this circuit, but only $P$ knows a set of assignments $(s_1, s_2, s_3, s_4, s_5)$ for the wires that satisfies each gate's rank-$1$ constraint. $P$ would like to reduce proving that $s_1 \cdot s_2 = s_4$ and $s_3 \cdot s_4 = s_5$ into proving a statement about polynomial divisibility, as the latter can be done succinctly.

Label the two multiplication junctions in the circuit with labels $x = 1$ and $x = 2$, which we call our target points. For each junction $i$, define polynomials $L_i(x), R_i(x), O_i(x)$, corresponding to the left, right, and output wires of junction $i$, respectively. The key requirement is that these polynomials must equal $1$ at $x = i$, and $0$ for other values of $x$. This interpolation property will be useful soon. To this end, let
\begin{align}
L_1 (x) = R_1(x) = O_1(x) &= 2 - x \\
L_2 (x) = R_2(x) = O_2(x) &= x - 1
\end{align}

Notice that $L_1, R_1, O_1$ are $1$ at $x = 1$ and $0$ at $x = 2$, and the opposite interpolation for $L_2, R_2, O_2$. Label the wires $j \in \{ 1,2,3,4,5\}$ and define $l(i) = j$ if $j$ is left wire into junction $i$; that is, its left wire's assignment is $s_j$. Define $r(i) = j$ if $j$ is right wire into junction $i$. Define $o(i) = j$ if $j$ is output wire of junction $i$. Now consider the polynomial,

\begin{align}
P(x) &:= (\sum_{i = 1}^{2} s_{l(i)} L_i) \cdot (\sum_{i = 1}^{2} s_{r(i)} R_i) - \sum_{i = 1}^{2} s_{o(i)} O_i\\
 &= \Big(s_1 (2 - x) + s_4 (x - 1)\Big) \cdot \Big( s_2 (2 - x) + s_3 (x - 1) \Big) - \Big( s_4 (2 - x) + s_5 (x - 1) \Big)
\end{align}

The interpolation properties becomes useful because requiring that $P$ have zeros at all target points (at $x = 1, 2$) yields the circuit's rank-$1$ constraints.
\begin{align}
P(1) &= s_1 \cdot s_2 - s_4 = 0\\
P(2) &= s_4 \cdot s_3 - s_5 = 0
\end{align}

Therefore, proving that $P(1) = P(2) = 0$ would imply that the rank-$1$ constraint at each gate is satisfied by the assignment. Proving that the polynomial $t(x) = (x - 1) (x - 2)$ divides $P(x)$ is equivalent to proving $P(1) = P(2) = 0$, which importantly offers a succinct way for the prover to convince the verifier that it possesses a valid assignment for the QAP. The crux of Pinocchio and Groth-16 is to prove this polynomial divisbility property holds without revealing the witnesses.

\begin{figure}[H]
\centering
\includegraphics[width=10cm]{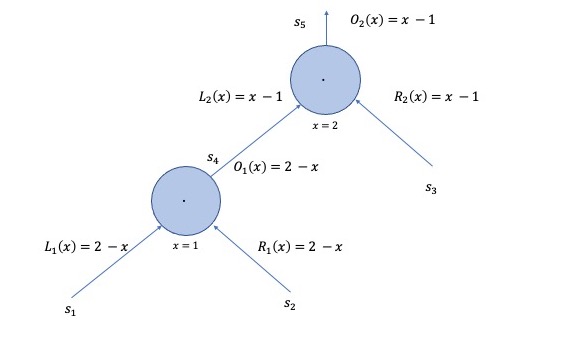}
\caption{Circuit from Figure 1, annotated with target points and polynomial interpolates}
\label{figure:arith-circuit}
\end{figure}

\subsubsection{The General Reduction}

The following reduction is adapted from \cite{vitalik-qap}.

\begin{enumerate}
    \item Suppose our circuit has $n$ wires and $m$ gates. Our witness is $s = (s_1, s_2, ..., s_n) \in \mathbb{R}^n$.
    
    \item For all $1 \leq i \leq m$ and $a_i,b_i,c_i \in \mathbb{R}^n$, a triple $(a_i,b_i,c_i)$ represents a rank-$1$-constraint for $s$, requiring that $(s \cdot a_i) \times (s \cdot b_i) - s\cdot c_i = 0$. There is one such constraint for each multiplication junction of our circuit.
    
    \item Stack the $m$ constraints vectors to obtain matrices $A, B, C \in \mathbb{R}^{m \times n}$, where 
    
    \begin{align}
    A = \begin{pmatrix}
a_1 ^T\\
a_2 ^T\\
\vdots \\
a_m ^T\\
\end{pmatrix}
B = \begin{pmatrix}
b_1 ^T\\
b_2 ^T\\
\vdots \\
b_m ^T\\
\end{pmatrix}
C = \begin{pmatrix}
c_1 ^T\\
c_2 ^T\\
\vdots \\
c_m ^T\\
\end{pmatrix}
    \end{align}

\item Using Langrange Interpolation \cite{lagrangeinterpolation}, find $3$ sets of $n$ polynomials $\{ u_i\}_{i = 1}^{n}, \{ v_i\}_{i = 1}^{n}, \{ w_i\}_{i = 1}^{n}$ such that for all $i \in [n]$ and $x \in [m]$,

\begin{align}
u_i (x) &= A[x][i]\\
v_i (x) &= B[x][i]\\
w_i (x) &= C[x][i]
\end{align}

\item Find the polynomial $h(x)$ such that $\Big( \sum_{i = 1}^{n} s_i u_i(x)\Big)\cdot \Big( \sum_{i = 1}^{n} s_i v_i(x)\Big) - \Big( \sum_{i = 1}^{n} s_i w_i(x)\Big) = h(x) t(x)$ \label{eq:qapred}

where $t(x) = (x - 1)(x - 2) ... (x - n)$.

The reduction is complete. Now $P$ can apply the Pinocchio protocol to prove to $V$ that she knows a witness $s$ that satisfies (\ref{eq:qapred}), without revealing her witness.

\end{enumerate}

\subsection{Pinocchio Protocol for QAP} 

\subsubsection{Non-zero-knowledge Pinocchio Protocol Construction}

We start by illustrating the workings for the Pinocchio protocol, without zero knowledge. Afterwards, we will state the modification to make the protocol zero knowledge. 

Let $G$ be a group of prime order $p$. Let $E : \mathbb{F}_p \rightarrow G$ be a homomorphic encoding. In particular, $E(x) = g^x$, for a generator g. Let $e : G \times G \rightarrow G_T$ be a elliptic curve bilinear map. Suppose the prover P knows a witness $\{s_i\}_{i = 1}^{n}$ for the original arithmetic circuit. By the reduction, she knows some polynomials such that 

\begin{align}
\Big( \sum_{i = 1}^{n} s_i u_i(x)\Big)\cdot \Big( \sum_{i = 1}^{n} s_i v_i(x)\Big) - \Big( \sum_{i = 1}^{n} s_i w_i(x)\Big) = h(x) t(x) \label{eq:verif}
\end{align}

The main idea of the protocol is as follows. $V$ would like to test $P$ at a random point $z \in \mathbb{F}_p$ for the values of the polynomials above. $P$ is queried for the value of the $u(z) = \sum_{i = 1}^{n} s_i u_i(z),  v(z) = \sum_{i = 1}^{n} s_i v_i(z), w(z) = \sum_{i = 1}^{n} s_i w_i(z)$, and $h(z)$ at some random $z \in \mathbb{F}_p$. $P$ will homomorphically encode these values and send them to $V$. Due to homomorphic and bilinear map properties, $V$ can verify that the homomorphically encoded values satisfy the same equation (\ref{eq:verif}). If they do, she can be confident $P$ truly knows a witness, without learning the witness itself. Below are more details of the main idea just described.

The first step of the protocol is to generate a common reference string (CRS), which contains homomorphic encodings of specific multiples of $z$. The CRS serves two purposes: First, $P$ can generate her homomorphic encodings for her proof using linear combinations of the group elements of the CRS without needing to know $z$. Second, setting up the CRS obviates the need for $V$ to manually generate a $z$, and send a message with the encodings of $z$ to $P$. This enables this proof to be purely non-interactive, as after the CRS is generated, $P$ has enough to generate a convincing proof. 

We can think of the CRS as two sets of public group elements: the evaluation key, which contains the group elements needed to construct the proof, and the verification key, which contains the elements needed to verify.

\textbf{Common Reference String:}

Pick random $\alpha, \beta_u, \beta_v, \beta_w, \gamma, z \in \mathbb{F}_p ^*$. Publish the CRS below, then discard all of the sampled group elements used in its generation (\textit{the toxic waste}).

\begin{multicols}{2}
Evaluation key: 
\begin{itemize}
\item $\{E(z^i)\}_{i = 0}^{n}$, $\{E(\alpha z^i)\}_{i = 0}^{n}$
\item $\{E(u_i(z))\}_{i = 1}^{n}$, $\{E(\alpha u_i(z))\}_{i = 1}^{n}$
\item $\{E(v_i(z))\}_{i = 1}^{n}$, $\{E(\alpha v_i(z))\}_{i = 1}^{n}$
\item $\{E(w_i(z))\}_{i = 1}^{n}$, $\{E(\alpha w_i(z))\}_{i = 1}^{n}$
\item $\{E(\beta_u u_i(z) + \beta_v v_i(z) + \beta_w w_i(z))\}_{i = 1}^{n}$
\end{itemize}

\columnbreak

Verification key: 
\begin{itemize}
\item $E(1), E(\alpha), E(t(z))$
\item $E(\gamma), E(\beta_u \gamma), E(\beta_v \gamma), E(\beta_w \gamma)$
\end{itemize}
\end{multicols}

\textbf{Prover's message:}\label{prover-msg}

Assuming the prover has polynomials $\Big( \sum_{i = 1}^{n} s_i u_i(x)\Big)\cdot \Big( \sum_{i = 1}^{n} s_i v_i(x)\Big) - \Big( \sum_{i = 1}^{n} s_i w_i(x)\Big) = h(x) t(x)$, first the prover computes $h(x)$. Then the prover's proof consists of the following:

\begin{itemize}
    \item $E(u(z)), E(\alpha u(z))$, where $u(z) = \sum_{i = 1}^{n} s_i u_i (z)$
    
    \item $E(v(z)), E(\alpha v(z))$, where $v(z) = \sum_{i = 1}^{n} s_i v_i (z)$
    
    \item $E(w(z)), E(\alpha w(z))$, where $w(z) = \sum_{i = 1}^{n} s_i w_i (z)$
    
    \item $E(h(z)), E(\alpha h(z))$
    
    \item $E(\beta_u u(z) + \beta_v v(z) + \beta_w w(z))$
    
\end{itemize}

\textbf{Verification:}

Upon receiving the prover's proof, the verifier first checks that the terms $u(z), v(z), w(z), h(z)$ were constructed as linear combinations of terms in the CRS, performing the following 4 checks.
\begin{align}
e(E(u(z)), E(\alpha)) &= e(E(\alpha u(z)), E(1))\label{eq:check1}\\
e(E(v(z)), E(\alpha)) &= e(E(\alpha v(z)), E(1))\label{eq:check2}\\
e(E(w(z)), E(\alpha)) &= e(E(\alpha w(z)), E(1))\label{eq:check3}\\
e(E(h(z)), E(\alpha)) &= e(E(\alpha h(z)), E(1))\label{eq:check44}
\end{align}


Next, check that each term $u(z), v(z), w(z)$ was generated using the same linear coefficients, $\{ s_i \}_{i = 1}^{n}$. That is, if $u(z) = \sum_{i} s_i u_i (z)$, then $v(z) = \sum_{i} s_i v_i (z)$ and $w(z) = \sum_{i} s_i w_i (z)$. This can be checked by verifying the following equation:

\begin{align}
e(E(\beta_u u(z) + \beta_v v(z) + \beta_w w(z)), E(\gamma)) = e(E(u(z)), E(\beta_u \gamma)) e(E(v(z)), E(\beta_v \gamma)) e(E(w(z)), E(\beta_w \gamma))\label{eq:check4}
\end{align}

Finally, we check the key condition that characterizes the polynomial divisibility criterion:

\begin{align}
e(E(u(z)), E(v(z))) = e(E(w(z)), E(1)) \cdot e(E(t(z)), E(h(z)))\label{eq:check5}
\end{align}

Accept if and only if all the checks (\ref{eq:check1})-(\ref{eq:check5}) hold.


\subsubsection{Non-zero-knowledge Pinocchio Protocol Analysis}

We will provide intuition for why this protocol works, deferring formal argument to the Pinocchio paper\cite{PHGR}. To begin, it is crucial to understand the following cryptographic assumption in order to understand the role the random offsets $\alpha, \beta_u, \beta_v, \beta_w$.

\begin{claim} (Knowledge of Exponent Assumption \cite{vitalik-snark})
Suppose Alice is given a pair of group elements $(x, \alpha x)$. Let's call such a pair $\alpha-$separated. Then it is computationally intractable for her to come up with another $\alpha-$separated pair $(y, \alpha y)$, except by deriving it as follows
\end{claim}

\begin{align}
(y, \alpha y) = (\gamma x, \gamma \alpha x)
\end{align}

An extension to this is that given $n$ $\alpha-$separated pairs, if Alice returns a different $\alpha-$separated pair, it must be a linear combination of the original $\alpha-$separated pairs with high probability \cite{vitalik-snark}.

Back to our protocol, the key condition that $V$ needs to check is the polynomial-divisibility condition: 
\begin{align}
u(z) v(z) - w(z) = t(z) h(z)
\end{align}

However, in addition to this equation, there need to be other equations that the verifier also checks. This is because it is possible for $P$ to forge values that satisfy this equation, but that were not generated from a true witness $s$ for the circuit. To this end, $V$ needs a way to check that the polynomials $P$ uses are truly a linear combination of the basis polynomials in the CRS.

The Knowledge-of-Exponent assumption is useful because if a pair of group elements that $P$ sends and a given pair are both $\alpha-$ separated, $P$'s pair must be generated as linear combination of the given $\alpha-$ separated pairs. Thus, $V$ can use a pairing function to efficiently check that these two pairs are both $\alpha-$ separated, to ensure $P$ produced her proof from an actual satisfying assignment of the circuit. More concretely, for two pairs $(x, \alpha x), (y, \alpha y)$, the following is true:

\begin{align}
E(x) &= g^x\\
e(E(x), E(\alpha y)) &= e(g,g)^{x \alpha y} = e(E(\alpha x), E(y))
\end{align}

$V$ can use this to produce the following check:

\begin{enumerate}
    \item  $e(E(u(z)), E(\alpha)) = e(E(\alpha u(z)), E(1))$ $\implies$ $u(z)$ is a linear combination of $\{u_i (z)\}_{i = 1}^{n}$

    \item $e(E(u(z)), E(\alpha)) \neq e(E(\alpha u(z)), E(1))$ $\implies$ $u(z)$ is \textit{not} a linear combination of $\{u_i (z)\}_{i = 1}^{n}$
\end{enumerate}

Using this idea, the verifier needs to verify three things

\begin{enumerate}
    \item $u(z) v(z) - w(z) = t(z) h(z)$
    
    \item $u(z)$ (resp. $v(z)$, $w(z)$) is a linear combination of $\{u_i (z)\}_{i = 1}^{n}$ (resp. $\{v_i (z)\}_{i = 1}^{n}$, $\{w_i (z)\}_{i = 1}^{n}$).
    
    \item $u(z)$ is a linear combination with the same coefficients as $v(z)$ and $w(z)$ (i.e. produced in linear combinations using the same witness $s$).
\end{enumerate}

We are now ready to describe why Pinocchio satisfies completeness and soundness.

\textbf{Completeness}

Regarding completeness, we are always able to generate such a proof from QAP. It remains to show that if $P$'s polynomials satisfy 

\begin{align}
\Big( \sum_{i = 1}^{n} s_i u_i(x)\Big)\cdot \Big( \sum_{i = 1}^{n} s_i v_i(x)\Big) - \Big( \sum_{i = 1}^{n} s_i w_i(x)\Big) = h(x) t(x)
\end{align}

then the verification checks (\ref{eq:check1})-(\ref{eq:check5}) hold. That is, the proof is valid and will be accepted with probability 1.

Equations (\ref{eq:check1})-(\ref{eq:check44}) hold if P honestly produces her polynomials from the CRS basis polynomials, due to the knowledge-of-exponent assumption.

Equations (\ref{eq:check4}) holds if P uses the same $\{ s_i\}_i$ as linear coefficients for her polynomials. This is evident as:

\begin{align}
e(E(\beta_u u(z) + \beta_v v(z) + \beta_w w(z)), E(\gamma)) &= e(E(\beta_u u(z)), E(\gamma)) e(E(\beta_v v(z)), E(\gamma)) e(E(\beta_w w(z)), E(\gamma))\\
 &= e(E(u(z)), E(\beta_u \gamma)) e(E(\beta_v v(z)), E(\gamma)) e(E(\beta_w w(z)), E(\gamma))\\
&= e(E(u(z)), E(\beta_u \gamma)) e(E(v(z)), E(\beta_v \gamma)) e(E(w(z)), E(\beta_w \gamma))
\end{align}

Equations (\ref{eq:check5}) holds due to the divisibility condition, Equation (\ref{eq:qap})

\bigskip

\textbf{Soundness}

Regarding soundness, if $P$ does not have a valid witness, then when $P$ constructs polynomials $u$, $v$, $w$, and $h$, then $u(x) v(x) - w(x) \neq h(x) t(x)$. Thus, it is enough to argue that if $P$ does not have polynomials satisfying Equation (\ref{eq:qap}), then $V$ will reject with high probability.

Suppose $u(x) v(x) - w(x) \neq h(x) t(x)$. First, the degree of $u$ and $v$ is $m - 1$, since they are the sum of polynomials who, constructed through Lagrange interpolation, go through $m$ pre-determined points (one for each junction in the arithmetic circuit). The degree of $t(x)$ is $m$. Therefore, the degree of $h(x)$ is $m - 2$. Then, $u(x) v(x) - w(x)$ and $h(x)t(x)$ are two polynomials degree $2m$. By the Schwartz-Zippel Lemma \cite{pinocchio-schwartz-zippel}, two different polynomials degree at most $2m$ intersect in at most $2m$ points. Thus, $u(x) v(x) - w(x)$ and $h(x)t(x)$ intersect in at most $2m$ points. Because $z \in \mathbb{F}_p$ has $p$ possible values, then with probability at least $1 - \frac{2m}{p}$, equation (\ref{eq:check5}) will fail and $V$ will reject $P$'s proof. This probability can be made arbitrarily close to $1$ by taking $p$ to be large.

\textbf{Proof and Verification Complexity}

The proof size is $9$ group members, from section (\ref{prover-msg}). Regarding verifier complexity, the verifier spends $8$ pairings to verify equations (\ref{eq:check1})-(\ref{eq:check44}). It spends $4$ pairings verifying (\ref{eq:check4}), and $3$ pairings for (\ref{eq:check5}). In total, this implementation of Pinocchio uses $15$ pairings. The CRS setup and the proof generation take time linear in the size of the original computation \cite{PHGR}.

Finally, the Pinocchio paper presents refined protocols that use $11$ pairings with a proof size of $8$ group elements \cite{Groth2016}.

\subsubsection{Zero-knowledge Pinocchio Protocol Modification}

In principle, if the verifier $V$ came up with their own witness $s' = (s_1', s_2', ..., s_n')$, they can compute $E(u'(z)), E(v'(z)), E(w'(z)), E(h'(z))$ following the prover's protocol. If these values are different from $E(u(z)), E(v(z)), E(w(z)), E(h(z))$, which the prover computed using $s$, then the verifier concludes that the prover's witness is not $s'$ \cite{pinocchio-schwartz-zippel}.

To eliminate this zero-knowledge violation, the prover $P$ adds on a random shift to the polynomials $u$, $v$, and $w$ \cite{PHGR}. The random shift will be a multiple of $t(x)$, so that everything is still the same $mod$ $t(x)$. For random $ \delta_1, \delta_2, \delta_3 \in \mathbb{F}_p$,
\begin{align}
u_z(x) &= u(x) + \delta_1 t(x)\\
v_z(x) &= v(x) + \delta_2 t(x)\\
w_z(x) &= w(x) + \delta_3 t(x)
\end{align}

Then, $P$ will evaluate these at $z$ using the CRS, and send over the shifted terms in its proof in the place of the corresponding unshifted terms.

\subsection{Groth-16 Protocol for QAP}

Groth-$16$ is a recent, more succinct zk-SNARK protocol that has replaced Pinocchio in many applications such as Zcash and Circom. It works on similar principles to Pinocchio, but both the proof size and the verifier complexity is significantly lower in Groth-$16$.

For an arithmetic circuit, let $n$ be the number of wires, $m$ be the number of gates, $P$ denote a pairing, and $E$ denote an exponentiation. Table (\ref{table:groth-stats}), taken from \cite{Groth2016}, compares the complexity of important quantities between Groth-$16$ and Pinocchio.

\begin{center}
\captionof{table}{\label{table:groth-stats}Comparison between Groth-$16$ and Pinocchio \cite{Groth2016}}
\begin{tabular}{ c |c| c| c| c}
  & CRS size & Proof size & Prover complexity & Verifier complexity \\ 
 Pinocchio & $7n + m$  $G$ & $8$ $G$ & $7n + m$  $E$ &  $11$ $P$\\  
 Groth-$16$ & $n + 2m$ $G$ & $3$ $G$ & $n + 3m$ $E$ &  $3$ $P $
\end{tabular}
\end{center}

\section{Application: Financial Security}\label{section:financeApps}

One of the most common use-cases for zero-knowledge proofs is for increased financial security. Although blockchains are already secure and decentralized, we can make even stronger security guarantees using zero-knowledge proofs, which allow us to, for instance, hide transaction amounts. In this section, we discuss two of the most common applications that deploy zero-knowledge proofs for financial security: Zcash and Tornado Cash.

\subsection{Zcash}

Zcash is a Layer $1$ (L$1$) token that allows for privacy protection \cite{zcash}. Zcash allows both private and public addresses, as well as both private and public coins. Coins can be ``shielded" or ``unshielded" by sending them from public to private or private to public addresses. In the following sections, we will examine a simplified form of Zcash with the strongest security guarantees -- a Bitcoin-like private ledger where coins cannot be traced back to the sender, the amount of each coin is shielded, and coins can be withdrawn from the ledger. \cite{zcashgithub, anatomyzcash}

To create a private transaction, Zcash uses a similar approach as Bitcoin, which validates the transactions by linking the sender address, receiver address, and the transaction's input and output values on the public blockchain. However, each of these values are shielded. Intuitively, one can imagine a naive implementation of Zcash that copies Bitcoin's unspent transaction output (UTXO) model, but where each UTXO's data is kept off-chain. In its place, a hash is kept on-chain, and a zero-knowledge proof is used to prove that some address has the right to consume the UTXO. We will construct the Zcash model piece by piece. 

The explanations in this section are largely derived from the Zcash whitepaper \cite{zcashpaper}.

\subsubsection{Protocol Construction}

The most naive model for Zcash would be as essentially an unordered set of commitments. To mint a coin, we sample some random $r_i$ and publish $c_i = h(r_i)$ for some hash function $h$. To allow Bob to spend $c_i$, we simply have to send Bob $r_i$ off-chain, and Bob can provide a zero-knowledge proof that his $r_i$ hashes to $c_i$, allowing him to consume the coin. The obvious difficulty here is that Alice still retains knowledge of $r_i$ through this process. The remainder of this section formalizes and elaborates on this basic idea.

{\bf Step 1: Anonymity with Fixed-Value Coins}. Let's begin with a model where all coins are discrete and have the same value. Each coin has a public serial number $sn$, together with with a secret randomness $r$. We then define a coin commitment as $cm \defeq \textbf{comm}_r(sn)$. A coin is then defined as the tuple $c \defeq (r, sn, cm)$. $\textbf{comm}$ is a one-way function such that $sn$ cannot be recovered from $cm$.

Now, consider the scenario where Alice wants to send a coin to Bob in a private transaction. To do this, the sender Alice submits the coin $cm$, and sends a corresponding coin to an escrow pool (e.g. $1$ BTC).

To withdraw the coin, Bob will reveal $sn$, together with a zero-knowledge proof of the following statement: ``I know $r$ such that $\textbf{comm}_r(sn) = cm$." If the proof is valid, then the 1 BTC gets withdrawn from the escrow pool and sent to Bob's address. $sn$ gets added to a list of revealed coins, so that no future spender may withdraw the same coin. So, for Alice to send the coin to Bob, she simply needs to send him $r, sn$ privately. 

Note that the original sender is anonymous. Although Alice's address is associated with a transaction containing $cm$, in order to associate $cm$ with $sn$ one would have to invert $\textbf{comm}_r(sn)$, which is assumed to be infeasible.

\textbf{Step 2: Problems with Naive Implementation.}

However, this implementation has many flaws. For instance, although we checked that $c$ wasn't spent before, we do not check that the $c$ belongs to Alice. In this model, once Alice puts a coin into the escrow pool, its original ownership is essentially erased, and anyone who knows $r, sn$ can withdraw the coin. This has a secondary problem, which is that when Alice sends the coin's information to Bob, Alice still retains the ability to spend the coin. One way to fix this is to transfer coins in pairs of transactions -- Bob spends coin $c$ and immediately mints $c'$ in the same block to protect himself. But this implementation is unwieldy.

This method also cannot be extended to a Bitcoin-like general-purpose ledger, since the coins are all fixed-size. Sending $100$ BTC would require minting 100 of these transactions, which is also unwieldy. Moreover, it is impossible to create coins with value of less than $1$ BTC. Both of these issues will be resolved with further uses of zero-knowledge proofs.

\textbf{Step 3: General-Purpose Anonymous Payments.}

To implement all of this, we redefine our commitments as follows. For a seed $x$, we define deterministic pseudo-random functions $\textbf{prf}^{addr}_x$ to generate public keys from secret keys, and $\textbf{prf}^{sn}_x$ to generate serial numbers. We define:

\begin{align}
pk = \textbf{prf}^{addr}_{sk}(0)\\
sn = \textbf{prf}^{sn}_{sk}(p)
\end{align}

Each user $u$ generates a public key and secret key pair $pk, sk$. The coins owned by $u$ contain $pk$ and can only be spent by proving knowledge of $sk$. $sk$ is randomly sampled, and $pk$ is related to it by $pk = \textbf{prf}^{addr}_{sk}(0)$. This allows us to generate a zero-knowledge proof relating $pk$ to $sk$.

To mint a coin of value $v$, the user first samples some randomness $p$, which is related to $sn$ as $sn := \textbf{prf}^{sn}_{sk}(p)$. Again, this allows for a zero-knowledge proof to verify that $sn$ was generated validly, without revealing $sk$ or $p$. $u$ then commits to $(pk, v, p)$ in two phases: (i) $u$ computes $k := \textbf{comm}_r(pk || p)$ for a random $r$; (ii) $u$ computes $cm: = \textbf{comm}_s(v || k)$ for a random $s$. This creates a nested commitment, which allows us to act on $v$ and $k$ without revealing the components of $k$. In particular, this allows a user to verify that a coin has value $v$ without actually revealing $pk$ or $sn$, which are a part of $k$.

We then define the \textbf{pour} function, which is a protocol the user follows to consume a bitcoin UTXO and is the main means by which our coins are exchanged. Suppose that $u$ with $(pk, sk)$ wants to consume $c = (pk, v, p, r, s, cm)$ and produce two new coins $c'_1, c'_2$ which satisfy $v'_1+v'_2=v$ targeted at addresses $pk'_1, pk'_2$. The user follows a two-step process for computing $c'_1, c'_2$. First, they sample $p'_1, p'_2$. This yields the coins $c'_1 = (pk'_1, v'_1, p'_1, r'_1, s'_1, cm'_1)$, $c'_2 = (pk'_2, v'_2, p'_2, r'_2, s'_2, cm'_2)$.

As part of the pour protocol, the user then provides a proof of each of the following four points to the verifier (the blockchain):

{\itshape
Given the serial number $sn$, and commitments $cm'_1, cm'_2$, I know $c, c'_1, c'_2$ and secret key $sk$ such that:

\begin{itemize}
    \item {The coins are well-formed: for $c$, $k = \textbf{comm}_r(pk || p)$, $cm = \textbf{comm}_s(v || k)$}, and similarly for $c'_1, c'_2$
    \item {The secret key and public key match: $pk = \textbf{prf}^{addr}_{sk}(0)$}
    \item {The serial number is well-formed: $sn = \textbf{prf}^{sn}_{sk}(p)$}
    \item {The values match: $v = v'_1 + v'_2$}
\end{itemize}
}

Upon executing this function, $u$ has mapped the value $v$ at some UTXO to two new UTXO's whose value add up to the original. This is the essence of a transaction. Thus, zero-knowledge proofs allow us to construct a decentralized payments protocol.

Note that the \textbf{pour} function also accomplishes a secondary purpose of allowing us to make coins harder to track. In particular, Zcash is backed using an escrow pool. This means that transactions can possibly be tracked by cross-referencing coin amounts to withdrawn amounts. Thus, \textbf{pour} also allows us to ``mix" the tokens. 

For this reason, in actual implementation we also modify \textbf{pour} to take two inputs and two outputs, which becomes a flexible general-purpose function that allows us to mix arbitrary inputs to produce arbitrary outputs, making them very difficult to trace. This allows us to more generally produce any number of inputs and any number of outputs, as we can also make input or output coins null. But for the purposes of this discussion, we will assume that \textbf{pour} only has one input.

\textbf{Step 4: Public Payments}

As described, the Zcash ledger can only be used for maintaining a private ledger of transactions. But this isn't very useful because in order to actually use any of these tokens, we need to withdraw coins from the shielded network at some point. Thus, to send coins to an unshielded address, we add an additional field $v_{pub}$ to \textbf{pour} which represents a public payment; the value-equation to prove now becomes $v = v'_1 + v'_2 + v_{pub}$. The address of this public payment is stored in transaction metadata, which is hashed on commitment and revealed on consumption. The public output is also optional, in which case $v_{pub}=0$.

\textbf{Step 5: Merkle trees}

Zcash provides one final innovation makes it more secure and scalable. Rather than directly proving knowledge of a coin, users will instead prove knowledge that this coin exists in the set of all commitments using a Merkle tree. The contract maintains a state root $rt$, and when we \textbf{pour} the coin $c$, we need to also provide a zero-knowledge proof of the Merkle proof that $rt$ contains $c$. By doing so, the Merkle tree is never revealed off-chain; it can be maintained by nodes or validators on the network.

The Merkle tree has an additional property of further masking transactions. It's possible by careful inspection to trace the history of a transaction. For instance, suppose that a coin is consumed and the entire value is made public. An attacker could trace the history of that coin and possibly see its source commitment. By putting this transaction data in a Merkle tree, we can entirely discard the commitments, meaning that at any given time, not only can adversaries inspecting the network not know the state of the ledger, they cannot even know its total size.

\subsubsection{Zcash Limitations}

Although Zcash is a novel and interesting concept, it still has many security risks. For instance, security is only offered at the level of the ledger. The nodes and the network can still be attacked. Even though public keys cannot be associated with private keys, an insecure IP address can still be associated with a public key. 

Additionally, since Zcash is a Layer 1 Bitcoin-like token, it has a hard time bootstrapping value into its own network. Unlike Tornado Cash (discussed in the next section), which is built on Ethereum, Zcash doesn't benefit from the battle-tested security of Ethereum. Users need to trust Zcash and its fairly centralized bridges and comparatively smaller network of validators. The security concept is novel but not clearly useful enough to demand an entirely new network. Indeed, only about 6\% of transactions on Zcash actually use the shielding functionality at all \cite{shieldedtx}, suggesting that for the most part, Zcash functions largely as yet another layer $1$ token.

Finally, though the use of Merkle trees also allows for further security on hiding transaction data, the implementation of them is difficult. In particular, the circuits get much more complicated, as the zero-knowledge proofs now also need to include Merkle proofs for both old coins. In addition, these Merkle proofs need to update along with the global state root. If a user waits a hundred years to unshield my Zcash coin, whatever original Merkle proofs they had are no longer valid. This means that there needs to be some sort of additional layer to maintain the Merkle trees, and the current implementation of Zcash simply ignores this by leaving the responsibility of tree maintenance to the nodes \cite{zcashpaper}.

\subsection{Tornado Cash}

Tornado Cash has a similar objective to Zcash, but rather than having private tokens that can be moved around and unshielded, only one transaction type is supported -- sending some amount of ETH untraceably from a sender to a receiver. However, on Tornado Cash, it's harder to trace the original transaction due to the use of a liquidity pool. The ``mixing" caused by this pool is what gives the Tornado protocol its name. (Zcash technically has some amount of mixing since each transaction mixes at least two inputs, but it's not as powerful as Tornado's single liquidity pool.) Zcash can be thought of as a bitcoin-like ledger where the state of the ledger is unknown but proofs of valid state transitions are published. However, each coin still has its own transaction history, and is in a sense non-fungible, since it has a serial number. On the other hand, Tornado Cash is much simpler, using a single giant liquidity pool that can only be deposited into and withdrawn from.

Specifically, we would like to be able to send an $N$ ETH note to an address $t$ without anyone to be able to trace where the transaction came from. In Zcash this is not quite possible, since we can think of Zcash as being essentially a shielded UTXO model. When each UTXO is consumed, the history of that consumption still exists somewhere on the blockchain -- it's just that the specific details of these actions are shielded. However, on Tornado Cash, the $N$-ETH funds are deposited into a liquidity pool on the contract. Users are able to view the total size of the liquidity pool, but not the size of the $N$-ETH notes that comprise it. This makes the transaction truly untraceable, as the original deposit transaction is not even stored anywhere on the contract.

The analysis in this section is largely drawn from the Tornado Cash whitepaper \cite{tornado}.  

\subsubsection{Protocol}

Formally, we allow a coin to be defined as $(k, r)$, where $k$ is a $n$-bit {\it nullifier} and $r$ is a $n$-bit {\it  randomness}; $k,r \in \mathbb{B}^{n}$ where $\mathbb{B} := \{0, 1\}$. Let $T$ represent a tuple of transaction data, chiefly including the designated address and the amount of ether sent. Let $H$ denote a hash function (MiMC, in our case).

When I {\bf deposit} a coin, I generate $k, r$ with my desired transaction $T$ and compute $C = H(k || r || T)$. I send $C$ to the contract, which stores it in a Merkle tree as a non-zero leaf node.

When I {\bf withdraw} a coin, I need to provide $k, r, T$, $C = H(k || r || t)$, together with a zero-knowledge proof that $C$ has been computed properly. Additionally, I need to provide a proof that $C$ is actually a member of the Merkle root currently stored in the contract. $k, T$ become public, while $r$ is kept private. In particular, $k$ is added to an array of nullifier hashes, which indicates which deposited coins have been withdrawn, so that no two coins can be withdrawn twice.

This protocol claims the following security guarantees:

\begin{itemize}
  \item Only coins deposited into the contract can be withdrawn
  \item No coin can be withdrawn twice
  \item Any coin can be withdrawn if $(k, r, T)$ are known
  \item Coins cannot be traced to their sender
\end{itemize}

\subsubsection{Tornado Cash Limitations}

There are still a few potential issues with the implementation. First of all, since the liquidity pool is entirely backed by these deposited notes, it's technically possible to track the source of a transaction simply by looking at changes in size of the liquidity pool, since users can track the total size of the pool on the contract. For instance, if the pool initially has $0$ ETH, one can track a single transaction that moves the liquidity pool from zero to a nonzero value, inferring that some address must initiate a transaction that moves the liquidity pool. This could possibly be resolved by having a large buy-in to the pool during setup phase, i.e. if the pool is initiated with $10000$ ETH, changes of $1$ ETH would be hard to track, since one could assume there would be many transactions of comparable size entering and leaving the pool, and moreover since these transactions would only be a fraction of the pool's total size.

However, very specific numbers could still be traced. For instance, one block causes the pool to increase by $3.267$ ETH, then a later withdraw transaction that decreases the pool by the same amount can be reasonably assumed to be linked. This could be resolved by some sort of transaction-mixing operation. For instance, a trusted 3rd-party relayer node could wait to receive a fixed size batch of transactions. That is, they could wait until $10$ ETH of transactions are submitted, then only make commitments of $10$ ETH at once while providing a proof that the sum of the transactions is $10$ ETH, perhaps supplying some of its own liquidity to round off. Similarly, a relayer could also take a large transaction and split it into smaller ones, e.g. a $50$ ETH transaction could be split into $5$ smaller ones of $10$ ETH each. 

Unfortunately, both of these changes require some higher-powered cryptographic tools. Abstractly, one wants to generate a SNARK that proves a valid state transition has occurred on the ledger, but which can encompass multiple transactions at once. This is certainly not impossible, and is in fact the same principle that rollups are built on, but it's not straightforward and detracts from the elegance of the Tornado Cash protocol.

Another issue is that transactions can be front-run. A front-run attack is when a node, who receives information from a client and is asked to execute a transaction, uses this information to their advantage. In this construction anyone who knows $(k, r, T)$ can force the transaction to occur. This differs from Zcash, where the user's address is actually taken into account in each transaction. Here, as long as the contract is provided with a valid proof, a coin can be released. While not typically an issue, since the recipient of the coin is encoded in the coin itself, this is still a potential security vulnerability in the sense that at an attacker is allowed to interfere with a process where neither party wants that attacker involved. One could imagine, for instance, if a large amount of funds are sent, that an attacker could force a transaction before the receiver sets up a secure way to receive the large amount of funds, such as a multisig or a hardware wallet.

\subsubsection{Comparison to Zcash}

Tornado Cash's key improvements on Zcash, are 

\begin{enumerate}
    \item The use of a liquidity pool to help mask transaction histories 
    \item The storage of some of the Merkle tree on the contract
    \item The move to a simple smart contract model rather than a complex layer-2 solution, which solves the issue of unmasking hidden coins when a user wants to withdraw them from privacy.
\end{enumerate}

It's worthwhile to compare these two solutions, which have similar end goals. The liquidity pool is certainly a very powerful innovation. On Zcash, it would be quite easy to trace a small number of tokens around just by following commitments manually, but with a liquidity pool, every transaction is truly untraceable. This, in addition to the fact that Zcash is built as a smart contract on top of Ethereum, which is an extremely well-trusted token with huge amounts of capital locked up, makes Tornado cash the platform of choice for most people, especially for sending large amounts of USD anonymously (since ETH is more popular than ZEC).

The use of a smart contract also means that consensus details can be handled by Ethereum, which allows the Tornado cash protocol to be much more elegant than Zcash's, as evidenced by the length of the protocol itself. This means that users don't need to trust smaller Zcash nodes but can instead place their trust in the Ethereum ecosytem as a whole.

However, the downside of this is that Tornado cash cannot entirely abstract the Merkle tree away from the protocol. Rather, it attempts to store some of the Merkle history in the contract. But this is not straightforward. In particular, it's not sufficient to simply store the root of the Merkle tree -- one also needs to store enough of a leaf path that the next leaf on the tree can be computed. This presents a possible security issue, since part of the guarantee Tornado cash makes is that transaction data is discarded, but in this case some amount of data needs to be retained so that the Merkle tree can be computed on the contract side. In particular, it stores the previous 100 Merkle roots \cite{tornado}, and a withdrawer can provide a Merkle proof to any of these Merkle roots. But since Merkle roots are recomputed each transaction, this means that a withdrawer needs to constantly update its Merkle proof: the withdrawer needs to provide a proof that the provided transaction is a leaf of the given Merkle root, so when the root changes, the proof changes. Then, if the withdrawer stops receiving state updates from the contract for 101 root updates, it will no longer have enough information to actually withdraw the coin, unless it receives the leaf path from a trusted 3rd party.

In any case, the use of a Merkle tree is actually subtly complicated here, which is the reason the nullifier is used -- rather than updating the Merkle tree to indicate when a coin has been consumed, it's easier to simply keep a list of all transactions that have ever been consumed. This seems rather inelegant; as this history will scale linearly with the number of coins ever consumed on the contract.

\section{Application: Rollups}\label{section:rollupApps}

In this section, we discuss ZK rollups. Rollups are tools to optimize blockchain transaction throughput by executing some portion of the transaction computation off-chain and using zero-knowledge proofs to verify these executions on-chain. Since it's faster to verify computations than execute them, this provides a significant speedup to the blockchain. Technically, a majority of current solutions for blockchains are only ``validity rollups," and don't exhibit perfect zero knowledge (\ref{appendix}). However, most of these rollups are implemented using libraries that do provide the zero-knowledge property, so we will call them ZK rollups.

These rollups are divided into two categories: constructions for application-specific rollups, where a particularly expensive part of an application is put into a rollup, and general rollups, which can generate validity proofs for any state transition on the EVM, scaling the entire Ethereum network. The former type of rollup is the kind that Zcash and Tornado Cash implement. In this section, we will start with a few more examples of application-specific rollups, like Dark Forest, and try to work towards a possible construction for general-purpose rollups, zkEVM.

\subsection{Validity Rollups and Validium}

First, we will introduce the intuition for rollups by discussing at a high-level how one might use general-purpose rollup to speed up Ethereum, assuming that one has a construction. Additionally, it's worth discussing the difference between two flavors of rollup. More broadly, one could categorize these as \textit{Validity-style} and \textit{Validium-style}, after the proposition for Validium as discussed in \cite{validium}. Both application-specific rollups and general rollups can be categorized into either of these types.

To begin, consider the problem we are trying to solve: although Ethereum is the most popular Turing-complete blockchain infrastructure on which people can build apps, the Ethereum block rate is merely $15$ second/block which can barely serve the billions of users it purports to serve. This leads to gas fees that often exceed hundreds of dollars.

For rollups, the key idea is that we allow users to compute transaction state changes off-chain. These computations are managed in a second network, called an L$2$, as opposed to direct on-chain (L$1$) computations. Moreover, there is a smart contract on L$1$ that links with L$2$, bridges tokens between the two, and verifies that L$2$ transactions happen correctly.

One way to implement this L$2$ scaling is called Optimistic Rollup, where we just assume that every L$2$ transaction is valid until someone refutes it. The contract holds for some period, giving users a chance to submit a fraud proof saying that a transaction was done incorrectly. We can see how optimistic rollup can scale Ethereum, but at the same time still relies on people submitting fraud proof to make sure that the transactions are secure.

ZK or Validity Rollup follows much the same architecture \cite{zkrolluparchitecture}, only instead of waiting for users to submit fraud proofs, Zero-Knowledge proofs are used to verify that transaction state changes are applied correctly \cite{vitalikrollup}. The general protocol is as follows:

\begin{itemize}
    \item {Users do their activities including signing transactions on L$2$ and submit those transactions to validators, which act as a bridge between L$1$ and L$2$}
    \item {Validators aggregate (``roll up") thousands of submitted transactions together into a single batch and submit the following to an L$1$ main chain smart contract:
    \begin{itemize}
        \item {A new L$2$ state root}
        \item{A zk-SNARKs to prove the correctness of the state root}
        \item{Transaction headers for each transaction in the batch}
        \item{A Merkle root of the transaction batch, allowing users to check whether or not a transaction was really in this batch}
    \end{itemize}
    }
    \item {The main chain, thanks to zk-SNARKs, verifies both the validity of all the transactions included in the block and the validity of the proposed state transition on the Merkle root}
\end{itemize}

zk-SNARKs verification is much more efficient than verifying every transaction individually. Also, storing the state off-chain is significantly cheaper than storing it on EVM, enabling boost of mainnet capacity and saving on transaction fees.

Validium essentially makes the minor change that even transaction headers are not stored, the SNARK instead proving that the state transition is valid, but not providing any information about what led to that state transition. We can think of validity rollups as being a proof of computation, and Validium as being both a proof of knowledge and of computation. The same differentiation exists for application-style rollups, too: one can roll up only an expensive computation but retain the data on-chain, or one can also discard the data and put it off-chain. We will call rollups as in the protocol described above \textit{Validity-style}, and protocols that also discard data \textit{Validium-style} \cite{validium}.

\subsection{Dark Forest}

Dark Forest \cite{darkforest} is the first fully on-chain game built on Ethereum, and is powered largely by zk-SNARKs. In the game, planet coordinates are kept private. Planet data is public, but is indexed by the hash of the planet's coordinates. The game is inspired by the highly adversarial nature of science-fiction novels but also of the Ethereum ecosystem more broadly. In particular, revealing one's coordinates can be seen as being akin to a declaration of war, since it invites other players to attack that planet.

Games, especially an MMORTS (massively multiplayer online real-time strategy games) like Dark Forest, are a very interesting use case for rollups.  In particular, due to the thousands of transactions that players execute when the game is in action, any amount of computation or storage that can be saved is incredibly valuable, since blocks will nearly always be entirely saturated. We will discuss two uses of rollups in Dark Forest.

In the game, each planet has cartesian coordinates $(x, y)$ and a planet ID $h := H(x, y)$ for some hash function $H$. Additionally, planets have certain properties, such as biome, computed from a Perlin noise function, denoted here as $n := p(x, y)$ for a Perlin function $p$ \cite{darkforest}.

A zk-SNARK is used to verify that $h = H(x, y)$. This not only protects the values of $(x, y)$, but also allows $(x, y)$ to not be stored on the contract. This, in a sense, is a Validium-style rollup. Perhaps more interestingly, Perlin noise can actually be quite expensive to compute on-chain, and it is computed off-chain in a Validium-style rollup.

The flow for generating a new planet thus is as follows: when a new planet is discovered, the user makes a transaction to the contract providing a proof that they know $(x, y)$ for some declared $h$ such that $h = H(x, y)$, and that the planet's Perlin value $n$ is also related to $(x, y)$ by $n = p(x, y)$. The contract verifies the first proof, and if it is successful, verifies the second proof. It then uses this value of $n$ and references it with a table of planet generation data, returning a planet with different properties depending on the value of $n$.

The full implementation of Perlin noise is not discussed here, but we will quickly give a construction for sampling a random gradient vector (one step in the Perlin noise process), since the example is illuminating. The typical way to do this would be by sampling a random angle and then calculating its sine and cosine. However, sine and cosine are not easy to linearize into an arithmetic circuit - they have Taylor series approximations, but those converge very slowly. The technique here is to use the hash of the coordinates to pseudorandomly generate an index from $0$ to $7$ as in $r(x, y)$ \% $8$, which is used to sample one of $8$ pre-generated vectors corresponding to $i\pi/4$ for $0 \leq i < 8$.

This construction shows how a specifically-engineered solution can provide a large speedup to a particular problem and reveals an obstacle for general-purpose rollups: generic algorithms for creating circuits are rarely as fast as specifically-engineered ones.

\subsection{zkEVM - Turing-Complete Rollups}

Here, we will discuss a particular implementation for general-purpose ZK rollups, as implemented by zkSync. This problem is often referred to as zkEVM, in the sense that we want to embed all state changes on the EVM into a circuit. We will begin with naive implementations, then describe the problems with them, moving towards a solution that is actually deployed onto Ethereum Mainnet \cite{zksync}.

We can essentially state the problem as such: for any given block, comprised of transactions that make calls to any number of contracts, we want to generate a zero-knowledge proof of the EVM state transition induced by the block.

\subsubsection{TinyRAM}

The most naive way to implement this is to first compile the block and its transactions down to its corresponding opcodes, then provide an arithmetic circuit node for each opcode. This is trivially easy to do, since this compilation must occur at some point, and since opcodes are read like a stack, it's easy to create a long circuit that simply reads opcodes off sequentially. zkSync in particular implements a TinyRAM architecture, which has a small instruction set, and thus can easily be converted into a circuit.

The problem with this implementation is that it's incredibly slow. Simple operations like array accessing might get compiled down to a very large number of opcodes -- incrementing pointers, etc. Because of this, TinyRAM has an average of over $1000x$ overhead compared to direct circuit implementation, like the one given above for Dark Forest \cite{zksync}.

\begin{center}
\captionof{table}{\label{table:tinyram}Comparison of circuit complexity for hardwired circuit vs TinyRAM \cite{zksync}}
\begin{tabular}{c | c| c}
 Operation & \# Gates: hardwired & \# Gates: TinyRAM \\
 Add & $1$ & $1$ K \\  
 Sub & $128$ & $1$ K \\
 Poseidon hash & $250$ & $>$ $250$ K \\
 SHA256 hash & $25$ K & $>$ $25$ M \\
 Keccak hash & $300$ K & unreasonably large \\
\end{tabular}
\end{center}

\subsubsection{Recursive Aggregation}

The next attempt would be to use a more powerful cryptographic tool called recursive SNARKs, discussed in section (\ref{recursive-SNARK}). Using elliptic-curve properties, it's possible to combine two SNARKs into one SNARK of the same size. The proposition here is for each contract to provide its own zero-knowledge proof, and thus a transaction would be able to aggregate a SNARK across each contract in a way similar to a Merkle tree.

For instance, consider a block comprised of 4 transactions: $t_1, t_2, t_3$ and $t_4$ together with proofs $\pi_1, \pi_2, \pi_3, \pi_4$. Recursive aggregation would allow us to generate $\pi_{12}$ from $\pi_1, \pi_2$ and $\pi_{34}$ from $\pi_3, \pi_4$, then finally $\pi_{block}$ from $\pi_{12}, \pi_{34}$. We continue this process for every transaction in a block to generate a proof for the whole block. Recursive aggregation is shown in the below figure:

\begin{figure}[H]
\centering
\includegraphics[width=8cm]{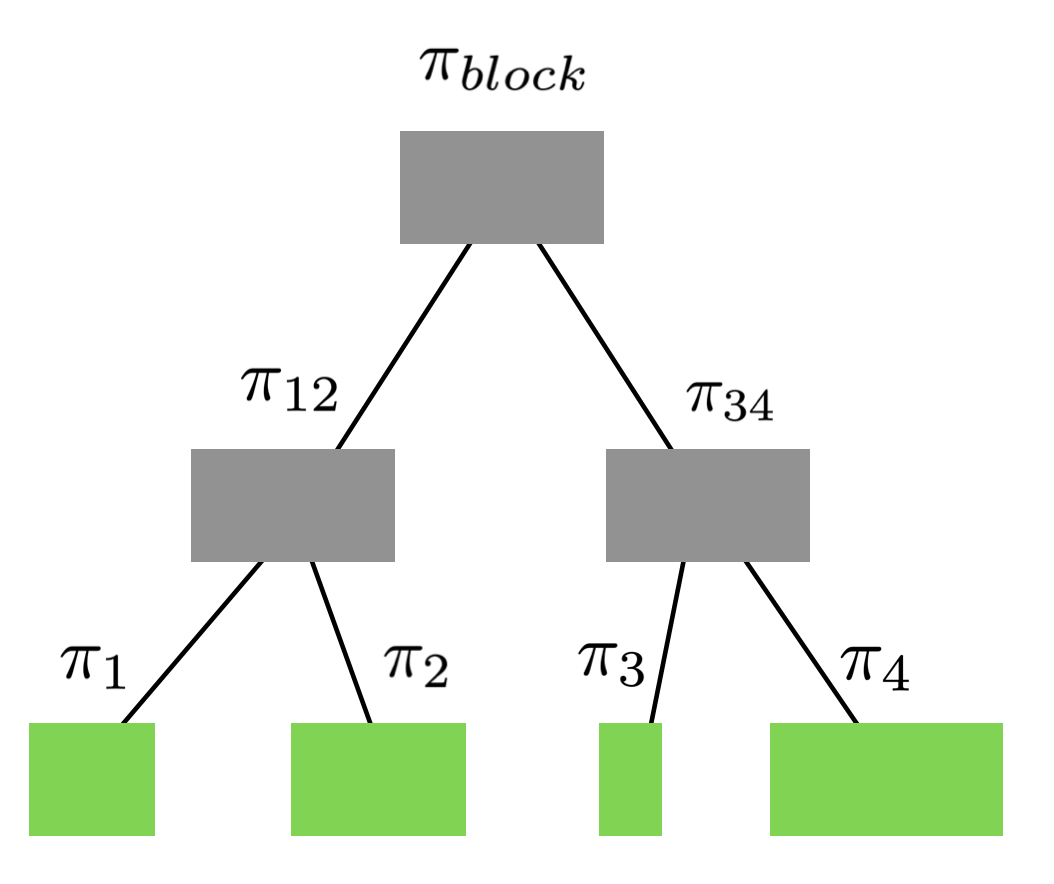}
\caption{Recursively aggregated SNARK}
\label{figure:aggregation}
\end{figure}

A crucial issue with this idea is that recursion-friendly cryptography isn't actually possible on the EVM--it's possible to implement, but there are no pre-compiles issued in the protocol. One possible solution is to use PLONK, a construction to generate a recursive SNARK, which was implemented by Matter Labs \cite{zksync} on January 2021. But the problem here is that PLONK is not Turing-complete. In particular, it cannot implement jumps or recursion, similar to a shader or other parallel operations.

\subsubsection{Heterogeneous Mixing}
The insight that allows zkSync to fully implement zkEVM is to use a mix of all three rollup solutions: hardwired circuits, TinyRAM, and recursively aggregated PLONK. This works because Ethereum opcodes tend to be distributed in cost as a power-law. That is, a small number of opcodes are more expensive than the rest combined: SSTORE costs $20000$ GWEI while ADD costs $3$ \cite{opcodes}. 

The result of this is that in most transactions, 99.9\% of the gas costs comes simply from memory and signatures. Thus, a naive solution to zkEVM is to write hardwired circuits just for the difficult operations (storage and signatures) while using TinyRAM for everything else. 

The problem with this naive solution is that circuits themselves also have a capacity. There's a maximum number of constraints that can be encoded into the arithmetic circuit, and these need to be allocated between these two solutions in some way. A naive fixed $50/50$ split doesn't work well because blocks tend to be fairly heterogenous. One block might comprise of many hashes, while the next comprises of SSTOREs. The result is that circuits will usually not end up being fully saturated, making the rollup inefficient.

The solution to this, and the solution that zkSync deploys for zkEVM, is \textit{hetereogenous mixing} -- the use of recursive aggregation to regulate the ratio between hardwired circuits and TinyRAM circuits. In essence, it combines all three of the discussed solutions for a maximally optimized rollup.

To see this in practice, suppose as in the previous example that we have $t_1, t_2, t_3$ and $t_4$ together with proofs $\pi_1, \pi_2, \pi_3, \pi_4$. We can think of each $\pi_i$ as being comprised of $\pi_i^t$, $\pi_i^h$, and $\pi_i^s$, referring to subroutines that deal with TinyRAM operations, hashes, and storage respectively (the three largest types of proofs). Recursive aggregation allows for each $\pi_i^t$ to be aggregated into one large $\pi^t$, and similarly for $\pi^h$ and $\pi^s$. $\pi_{block}$ is then aggregated from $\pi^t$, $\pi^h$, and $\pi^s$. This situation is illustrated in the below figure:

\begin{figure}[H]
\centering
\includegraphics[width=8cm]{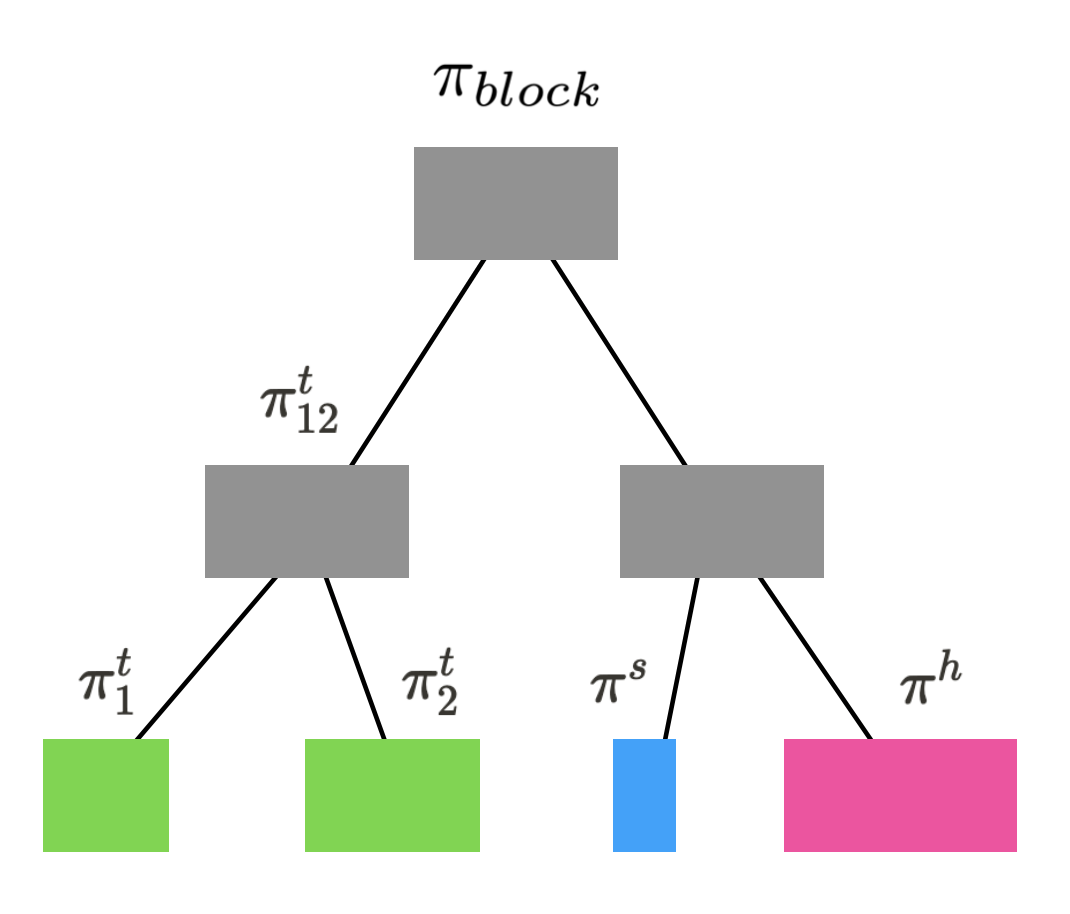}
\caption{Hetereogenous mixing for SNARKs}
\label{figure:aggregation}
\end{figure}

In summary, combining these various techniques allows us to generate Turing-complete zero-knowledge proofs. This, together with application-specific rollups, will allow Ethereum to scale significantly over the coming years.

\section{ZK Circuits for Novel Applications}\label{section:newApps}

This section provides novel circuit designs for two applications: private auctions and decentralized card games. These protocols use zk-SNARKs to provide unique guarantees that are necessary for these applications. Code for these designs can be found in the appendix (\ref{section:auctioncode}).

\subsection{Private Auction}

\subsubsection{Motivation}

Zero knowledge private auctions aim to simulate an anonymous auction process where buyers can submit bids without revealing their identity or funds available. Maintaining such private information is crucial in an auction. In a recent event involving ConstitutionDAO, a group of crypto holders collected over $40$ million dollars in ETH to bid for one of the earliest copies of US Constitution available. However, because their total funds were publicly viewable on the blockchain, another group in the auction outbid their maximum by a small amount and won the auction \cite{constitutionDAO}.

Creating a decentralized private auction is complicated for multiple reasons. First, the identity of the accounts need to be private because revealing the account reveals their available funds for the auction. Second, the auction needs to prevent malicious users from submitting bids beyond their available funds. There should be penalty if the user fails to pay the winning bid. 

A naive solution to these two problems is where the user pays their bid upfront for each bid they make. However, this creates many additional transactions and results in an overly complicated protocol. Any form of transaction between the money holding account and the auction contract reveals information about the bidder. By searching for the auction contract on Etherscan \cite{etherscan}, people can view all transactions and the originated addresses. Hence, the bidding process cannot involve a direct transfer of funds to the contract performing the auction.

Our proposed solution involves a proof-of-stake process where the bidders stake a certain amount of coins upfront, an auction entry fee. If the bidder did not win the auction, the coins can be reclaimed. If the bidder wins the auction, they needs to pay the remaining amount on top of the staked coins within a given time period to avoid the coins getting slashed. \\

\subsubsection{Accounts}

Each user should have two accounts. First, they have a staking account, which is public and is used to transfer the entry-fee coins up front. Additionally, the seller may choose to double the staking coins in the middle of the auction, which will require the bidder to transfer more coins to the auction from the staking account. The money transfer in between accounts can be untraceable with use of Tornado Cash.

Second, each bidder also has a private account. This account holds the actual funds the bidder will use to fulfill the winning bid. Because this account is separate from the staking account, a third party cannot determine the maximum a bidder can bid from the transaction from their staking account. Also, to protect the interests of the seller, zero knowledge cryptography should be used to prevent the bidder from submitting bids higher than the actual funds.

\subsubsection{Protocol}

The protocol should consist of the following steps.

\begin{enumerate}
    \item 
    
The users pre-commit the MiMC hash of the account address and the MiMC hash of funds available to the Merkle Tree. The zero knowledge proof will verify that the user indeed owns the account, and that the pre-images of the commited MiMC hashes are equal to the public account address and the available funds. Account addresses to the Merkle tree cannot be duplicated. 

    \item
    
    Before a bidder submits a bid, the contract verifies the previous bidder's account. The previous bidder cannot be the same as the current bidder, to prevent one account artificially inflating the price. Although this mechanism does not prevent a person from having two staking accounts and two private accounts, there is a deterrent to this behavior: having two accounts will reduce the fund available in each account and therefore reduce the possibility of winning the auction, as transfer of funds into the private account is not possible after commitment to Merkle tree.

    \item
    
    A zero knowledge proof will be submitted to verify that the bidder owns a fund that is higher than the bid. The input to the circom circuit includes 

\begin{enumerate}
    \item [(1)] pre-image of the account address 
    \item [(2)] pre-image of the available funds 
    \item [(3)] bid submitted to the contract. 
\end{enumerate}

A comparator circuit is used to compare the available funds with the bid submitted to the contract. If the comparator zero-knowledge circuit returned True, the bid is valid. Else, the bid is invalid. The circuit also consists intermediary checks to prevent tempering the witness file. The circuit also checks that the MiMC hash of account address and available funds is in the Merkle tree. As such, we can have a complete protocol of a private decentralized auction process. 
\end{enumerate}

There are possible attacks against this protocol. For example, a person can use alternative sources like Etherscan to look for accounts with most funds. MiMC hash is not reversible, but it will be easy to compute the MiMC hash of the top 1000 available funds. If all the 1000 funds are not part of the auction, this will give some upper bounds on the amount of funds available in the merkle tree accounts. If one or more of the funds are part of the auction, a lower bound on the amount of funds can also be calculated. With prior knowledge on the possible participating funds, it is easily verifiable if a given fund participate in the auction or not. 

One idea to prevent this attack is for the user to each to commit a salt -- a large random number that will be added together with the account address and fund available inputs. This will make the attack invalid because knowing the address itself does not generate the hash - the pre-image of the salt needs to be known too.

Our implementation of the above protocol can be found in the appendix (\ref{section:auctioncode}).

\subsection{Decentralized Card Games}

\subsubsection{Motivation}

We provide a set of protocols that can be used to play incomplete information games on chain (games where the players don't have full information of the game state). For example, in poker the player's hand is private information, and the card drawing process is completely private as well. Incomplete information is interesting because it allows for game theory, heuristic reasoning, bluffing, and other complex strategies, even outside the context of card games. For instance, given the recent popularity with non-fungible tokens (NFTs), it may be interesting to see players using NFTs as their ``cards" to play various games of incomplete information. 

However, incomplete information games are hard to simulate on chain. It is difficult to simultaneously prevent cheating and ensure privacy in games, as the hidden cards have to be stored and drawn locally without explicit commitment to the smart contract. 

The current card games use a commit and reveal approach to achieve privacy and anti-cheating. Players commit each action they take in a game by committing the hashes. For example, if a player claim that they have played the card $5$, then they must commit the hash of their remaining hand on chain. At the end of the game, all hands are revealed publicly and verified at each step to ensure that no players are lying in the game. If any discrepancies are found in the process, the entire game is rolled back to the beginning. 

The commit and reveal process is problematic for two reasons. Firstly, there is no way to verify that rules were followed before the end of the game. There is no way to catch players lying on the spot, and any slight discrepancies or misplay will have to void the entire game. The roll-back process is costly and frustrating, especially if players deliberately try to void the game. Secondly, the commit and reveal approach requires that all cards be shown publicly at the end, revealing each player's bluffing strategy. Bluffing is a key strategy in a poker game and information about players' strategies should not be revealed if it can be avoided.

An interesting question is whether we can leverage zero knowledge proofs to solve these two problems. For this, we provide the following decentralized card game protocols that don't have the two issues mentioned above.

\subsubsection{Decentralized Card Game Subroutines}

\textbf{Generating Randomness}

Randomness is required during the card draw process to get a card. Suppose the card deck is a public array with integers $1$ to $13$. Then, the following process ensures a random card is drawn. The following protocol can be used to generate public and private source of randomness. 

\begin{enumerate}
    \item A public source of randomness is given by previous block's hash. Whenever a new block is committed, its block hash can be assumed to be pseudo-random. However, relying solely on previous block hash can lead to bias. Players can choose when they play the card, depending on if the previous blockhash is would yield a favorable draw. 
    
    \item The private source of randomness is given by a seed hash committed by the player prior to the start of the game. The player is able to choose a random number $k$, and the MiMC hash, MiMC($k$), will be pre-committed on chain. During the game, the player is able to prove their knowledge of the pre-image of MiMC($k$) without revealing $k$ itself. The number $k$ constitutes private source of randomness.

\end{enumerate}

\textbf{Random and Private Card Draw}

With the ability to simulate randomness, the random and private card draw is fairly straight forward. A random number can be generated from previous block hash and player's pre-commited hash, possibly as an addition of the two numbers. The card drawn will be given as: 

\begin{align}
\textrm{card} = (\textrm{previous block hash} + k) \% (\textrm{number of cards})
\end{align}

To further reduce the bias of the game, it is also possible for each player to set the secret seed $k$ for their opponents. Such protocol is able to achieve unbiased random card draw while ensuring privacy.  

\textbf{Check Hand Consistency}

While it is straightforward to draw a card, the process where a hand is updated is not as obvious. Note that the player's hand is also private. When a card is drawn, the hand needs to be updated, and the hash of the new hand needs to be committed on chain. The zero knowledge protocol needs to achieve three purposes. 
\begin{enumerate}
    \item The protocol needs to ensure that the old hand corresponds to the previous committed hash. 
    \item It needs to prove transition between the old hand and new hand. 
    \item It needs to ensure that the committed hash corresponds to the hash of the new hand. 
\end{enumerate}

While the first and third steps are a simple implementation of MiMC module in Circom \cite{circom}, the second step is not as obvious, where we need to prove transition between old hand to new hand. With this challenge, a permutation protocol can be used where we prove the (old hand $+$ card drawn) is a permutation of the (new hand $+$ empty card). The card drawn process is seen as a exchange between an empty card from old hand with the new card drawn. 

\textbf{Private Card Play}

During private card play, a card can be publicly shown. The zero knowledge proof process is fairly similar to the previous steps. The player needs to go through three-stage proof process similar to the previous parts. Firstly, prove the old hand is the pre-image of the committed hash. Secondly, prove transition from old hand to new hand through permutation. (old hand + empty card) is the permutation of (new hand $+$ card played).

With these subroutines, we now have the necessary building blocks to generate many possible decentralized card games on chain. A proof of concept implementation can be found in the appendix (\ref{section:cardgamecode}).

\section{Future of ZK}\label{section:future}
\subsection{zk-STARKs}

Although success of zk-SNARKs is shown through various applications such as Zcash and zk-Rollups, they still have exploitable weaknesses. zk-SNARKs require a trusted setup, which means that it needs a certain secret key to create the common reference string, on which proof and verification is based. This private key is called ``toxic waste" because it needs to be disposed or securely kept. If an attacker obtains this secret, he can utilize the secret to forge the transaction \cite{starkvssnark}.

One common way to mitigate the risk of attacker acquiring the toxic waste is via Multi-Party Computation, where we require a set of multiple participants to cooperatively construct the key. Each of them holds a shard of private key which is for constructing a shard of public key (CRS) that can be combined to get public key \cite{toxicwaste}. In this setup, the exploit is minimized since having at least one of participants successfully deleting their private key shards is enough to make it impossible for an attacker to acquire toxic waste. However, this approach can be cumbersome. Another concern of SNARKs is that it is not quantum-computer resistant.

zk-STARKs was created by Eli Ben-Sasson, Iddo Bentov, Yinon Horeshy in 2018 \cite{binancestark}. Unlike zk-SNARKs, the  base technology for zk-STARKs relies on collision-resistant hash functions. As a result, zk-STARKs doesn't require an initial trusted setup and also achieve quantum-resistance. However, zk-STARKs proof has a far bigger size of the proofs compared to zk-SNARKs, resulting in much longer verification time than zk-SNARKs and costing more gas. Although the developer documentation, tools, and community of zk-STARKs is far smaller than zk-SNARKS, zk-STARKs is a promising, emerging technology, as seen from the fact that the Ethereum Foundation gave STARKware, a zk-STARKs based scaling solutions, a \$12 million grant.

\subsection{Recursive SNARKs} \label{recursive-SNARK}

Although SNARKs can be applied to various applications, there are some applications that naive SNARKs is not suitable. For example, in case that we want to prove the correctness of a function after $t$ iterated execution on it. With SNARKs, we need to prove all $t$ executions at once. This ``monolithic" execution can cause many problems. For instance we may not know number of $t$ in advance or the whole $t$ executions are too big for memory \cite{overviewrecursive}. This problem becomes clearer when we want to implement an application like private but verifiable elections. In these elections, people should be able to vote without exposing their identity, and the protocol can count the votes and verify that the vote result is correct. As we've discussed above, this application is almost impossible with naive SNARKs since we are unable to process all the proofs of each individual voter all at once. 

As a result, another version of SNARKs is developed, called ``recursive SNARKs." With recursive SNARKs, it is possible to apply SNARKs at each iterated execution to prove that execution and the correctness of prior proof. Hence, we don't need to wait to aggregate all executions to prove at once. With this, the election problem becomes a lot easier since people can submit the proof of their vote and the vote-counter can just aggregate the vote and verify the intermediary result of voting. When more people votes are sent, we can just apply recursive SNARKs again to update and verify the new voting result\cite{recursnarks}.

\newpage
\section{Acknowledgments}
The authors would like to thank Tim Roughgarden and Maryam Bahrani for their support in teaching the material necessary to write this paper.

\section{Appendix}\label{appendix}

\subsection{Formal Definitions}

This section provides formal definitions for soundness, completeness, zero-knowledge, touching on the difference between computational and perfect versions of these definitions. The following definitions are directly quoted from Groth, Ostrovsky, and Sahai's 2011 paper \cite{Groth2011}. 
 
 Let $R$ be an efficiently computable binary relation. For $(x,w) \in R$, we call $x$ the statement and $w$ the witness. $L$ is the language consisting of statements in $R$. A non-interactive proof system for relation $R$ consists of a common reference string generation algorithm $K$, a prover $P$, and a verifier $V$. These algorithms are all probabilistic non-uniform polynomial time algorithms. $K$ produces a CRS $\sigma$ of length $\Omega(k)$. The prover takes $(\sigma, x, w)$ and produces a proof $\pi$. The verifier takes $(\sigma, x, \pi)$ and outputs $1$ if and only if the proof is accepted. 
 
 Denote $1^k$ as a unary string of $1$'s of length $k$. Denote $\sigma \leftarrow K(1^k)$ as $\sigma$ being assigned the output of $K(1^k)$, with probability equal to the probability that $K$ outputs $\sigma$. $(K, P, V)$ is a non-interactive proof system for $R$ if it has completeness and soundness properties below:

\begin{definition} (Perfect Completeness) A proof system is complete if for all adversaries $A$

\begin{align}
P[\sigma \leftarrow K(1^k); (x, w) \leftarrow A(\sigma); \pi \leftarrow P(\sigma, x, w) : V(\sigma, x, \pi) = 1 \text{ if  } (x, w) \in R] = 1
\end{align}

\end{definition}

\begin{definition} (Perfect Soundness) A proof system is sound if for all polynomial size families $\{ x_k\}$ of statements $x_k \notin L$ and all adversaries $A$,

\begin{align}
P[\sigma \leftarrow K(1^k); \pi \leftarrow A(\sigma, x_k) : V(\sigma, x_k, \pi) = 1 ] = 0
\end{align}

\end{definition}

\begin{definition} (Computational Soundness) A proof system is computationally sound if for all polynomial size families $\{ x_k\}$ of statements $x_k \notin L$ and all non-uniform polynomial time adversaries $A$, if $k$ is sufficiently large,

\begin{align}
P[\sigma \leftarrow K(1^k); \pi \leftarrow A(\sigma, x_k) : V(\sigma, x_k, \pi) = 1] \leq k^{-c}, \forall c > 0
\end{align}

\end{definition}

\begin{definition}(Computational Zero Knowledge)

A non-interactive proof $(K, P, V)$ is computational zero-knowledge if 

\begin{enumerate}
    \item There exists a polynomial time simulator S = $(S_1, S_2)$ where $S_1$ returns a simulated CRS $\sigma$ with a simulation trapdoor $\tau$ that enables $S_2$ to simulate proofs without access to the witness. In particular, if $(x, w) \in R$, $S(\sigma, \tau, x, w) = S_2(\sigma, \tau, x) $. If $(x, w) \notin R$, both oracles output ``failure".
    
    \item For all non-uniform polynomial time adversaries $A$,

\end{enumerate}

\begin{align}\label{comp-zero-knowledge-formal}
P[\sigma \leftarrow K(1^k) : A^{P(\sigma, \cdot, \cdot)} (\sigma) = 1]\approx P[(\sigma, \tau) \leftarrow S_1(1^k) : A^{S(\sigma, \tau, \cdot, \cdot)} (\sigma) = 1]
\end{align}

\end{definition}

The notation $A^{P(\sigma, \cdot, \cdot)}$ and $A^{S(\sigma, \tau, \cdot, \cdot)}$ mean that $A$ has access to an oracle that on input $(x, w)$ returns a proof $\pi$. The adversary only sees the inputs and outputs of the oracle, and does not know which type of oracle it has access to. 

Intuitively, the adversary $A$ plays the role of a distinguisher between the distributions over proofs generated by the simulator and by the actual prover. The more perfectly some simulator can simulate the prover, the more the proof is zero knowledge, because then the prover does not reveal anything to the verifier that would help the verifier compute anything much faster than before \cite{Goldwasser1989}. Everything the verifier sees from the prover is something the prover could have computed for herself if she knew that $(x, w) \in R$.

Taking this idea further, if the two probabilities in (\ref{comp-zero-knowledge-formal}) are exactly equal, then $(K, P, V)$ is \textit{Perfect Zero Knowledge}.

There is an important distinction between the computational and perfect versions of soundness and zero-knowledge. Computational soundness is weaker  than perfect soundness, requiring that only non-uniform polynomial time adversaries cannot provide accepted proofs for false statements with non-negligible probability. Similarly, computational zero-knowledge is weaker than perfect zero-knowledge, requiring that the information the verifier sees, called its View \cite{Goldwasser1989}, is distributed similarly to that of some polynomial time simulator.

With these definitions, the distinction between proof systems and argument systems is as follows. Proof systems require perfect soundness––a computationally unbounded prover cannot make proofs for false statements. On the other hand, argument systems only require computational soundness. 

This distinction is made because different notions of zero-knowledge are achievable in argument systems versus in proof systems. Fortnow \cite{Fortnow1987} showed that unless \lang{PH} collapses, there do not exist perfect zero-knowledge proof systems for \lang{NP} complete problems. In fact, even interactive proof systems cannot have both perfect soundness and perfect zero knowledge. In contrast, there are perfect zero-knowledge (and additionally, non interactive) argument systems for \lang{NP} complete problems, namely the zk-SNARK protocols \cite{Groth-10, Groth2011}.

\subsection{Code for Novel Applications}

\subsubsection{Private Auctions}\label{section:auctioncode}

As a proof of concept, we implemented a minimalistic version of zero-knowledge private auction in Circom with accompanying contracts. The zero knowledge proof below verifies if a given bid is valid at each instance. The circuits verifies Merkle Tree identitiy inclusion for a given user, as well as proving bid validity. 

To prove the Merkle tree inclusion, the inputs to the circuit consists of (a) the leaf node, (b) the root node, (c) pathElements, (d) pathIndices. This information is either committed on chain or can be easily generated when user commits to Merkle tree. pathIndices consist of a list of $0$ or $1$ that indicates if a given pathElement is on the left or right of the Merkle Tree. By continuously hashing the current node with the pathElements, we are able move up in the Merkle Tree till the root node. Then, the inclusion check is valid if the calculate root node is equal to the public root node, suggesting that a path exists to move from the leaf node to the root node. This part of the circuit is generated with reference to Tornado Cash and Semaphore implementation of the Merkle tree. Since the use of MiMC hash function exceeds the number constraints given by Circom, Poseidon hash \cite{Grassi2020} is used as an alternative, a more zk-SNARKs-friendly version of the hash function.

To prove the validity of the bid, we need to ensure that the given bid is lower than the funds available. The inputs to the circuit consists of (a) pre-image of the account, (b) pre-image of the fund available, (c) submitted bid. The Poseidon hash of the pre-image of the account and pre-image of the fund available will be checked against the publicly committed hash, which in our case is the leaf node. If equal, we can verify that the bidder is indeed the owner of a participating account. Then, a comparator circuit is used to verify that the fund available is larger than the bid. This prevents overbidding to take place, which can hurt the decentralized auction ecosystem.

The output the circuit mainly consists of (a) outValid, which indicates if the available fund is higher than the bid, (b) finalBid, which is strictly equal to the bid submitted to the user. Note that the outputs of Circom is directly fed into the smart contract. By having finalBid as an output, this ensures that the smart contract only accepts the bids that goes through the zero knowledge proof circuit in Circom. 

\begin{lstlisting}

include "../../node_modules/circomlib/circuits/mimcsponge.circom";
include "../../node_modules/circomlib/circuits/poseidon.circom";
include "../../node_modules/circomlib/circuits/mux1.circom";
include "../../node_modules/circomlib/circuits/comparators.circom"

template PoseidonHashT3() {
    var nInputs = 2;
    signal input inputs[nInputs];
    signal output out;

    component hasher = Poseidon(nInputs);
    for (var i = 0; i < nInputs; i ++) {
        hasher.inputs[i] <== inputs[i];
    }
    out <== hasher.out;
}

// Computes PoseidonHash([left, right])
template HashLeftRight() {
    signal input left;
    signal input right;
    signal output hash;

    component hasher = PoseidonHashT3();
    hasher.inputs[0] <== left;
    hasher.inputs[1] <== right;
    hash <== hasher.out;
}

// if s == 0 returns [in[0], in[1]]
// if s == 1 returns [in[1], in[0]]
template DualMux() {
    signal input in[2];
    signal input s;
    signal output out[2];

    s * (1 - s) === 0
    out[0] <== (in[1] - in[0])*s + in[0];
    out[1] <== (in[0] - in[1])*s + in[1];
}


template BidVerifier(levels) {
    signal input leaf;
    signal input root;
    signal input pathElements[levels];
    signal input pathIndices[levels];
    
    signal private input account;
    signal private input value;
    signal private input bid;

    // Merkle tree inclusion check
    component selectors[levels];
    component hashers[levels];
    signal output hashP;
    signal output outValid;
    signal output finalBid;

    for (var i = 0; i < levels; i++) {
        selectors[i] = DualMux();
        selectors[i].in[0] <== i == 0 ? leaf : hashers[i - 1].hash;
        selectors[i].in[1] <== pathElements[i];
        selectors[i].s <== pathIndices[i];

        hashers[i] = HashLeftRight();
        hashers[i].left <== selectors[i].out[0];
        hashers[i].right <== selectors[i].out[1];
    }

    root === hashers[levels - 1].hash;

    // Bid validity check
    component hasher = PoseidonHashT3();
    hasher.inputs[0] <== account;
    hasher.inputs[1] <== value;
    hashP <== hasher.out;
    hashP === leaf;

    component greater = LessThan(11);
    greater.in[0] <== value;
    greater.in[1] <== bid;

    outValid <== greater.out; 
    finalBid <== bid;
}

component main = BidVerifier(16);
\end{lstlisting}

\subsubsection{Decentralized Card Games}\label{section:cardgamecode}

We implemented a simplified poker game to demonstrate the randomness simulation and card playing process. The simplified game rule consists of a player drawing a card and comparing the card with the dealer's card. The person with the larger card number wins the game.

There are two main circuits involved in the process. The first circuit is used to randomly and privately draw a card. It takes in the blockhash and private seed as inputs. Using the module circuit, it is able to divide (seed $+$ blockhash) by $13$ and uses the remainder as the card drawn. The output seedCommit can be used to verified against the previously committed hash on chain. The output cardCommit can be used to directly commit the new card hash on smart contract.

\begin{lstlisting}

include "../../node_modules/circomlib/circuits/mimcsponge.circom"
include "../../modulus.circom"

template Main() {
  signal private input seed;
  signal input blockhash;
  
  signal output cardCommit;
  signal output seedCommit;
  
  signal card;
  
  component cardCalculator = Modulo(16, 100000000000000000000000000000000000000);
  
  cardCalculator.dividend <== seed + blockhash;
  cardCalculator.divisor <== 13;

  card <== cardCalculator.remainder + 1;

  component cardHash = MiMCSponge(1, 220, 1);
  cardHash.ins[0] <== card;
  cardHash.k <== 0;
  cardCommit <== cardHash.outs[0]

  component seedHash = MiMCSponge(1, 220, 1);
  seedHash.ins[0] <== seed;
  seedHash.k <== 0;
  seedCommit <== seedHash.outs[0];
}

component main = Main();

\end{lstlisting}

The second circuit involves comparing the player's card with the dealer's card to determine the winner. MiMC hash of the playerCard is compared with the previously committed hash. A comparator circuit is used to determine if the playerCard is larger than the dealer's card. It is interesting to note that the player's card is a private input and remains hidden throughout the game. There is no reveal step during the entire decentralized card game.

\begin{lstlisting}
include "../../node_modules/circomlib/circuits/mimcsponge.circom"
include "../../node_modules/circomlib/circuits/comparators.circom"

template Main() {
  signal private input playerCard;
  signal input playerCardCommit;
  signal input dealerCard;

  signal output outCardCommit;
  signal output outValid;

  /*
    Verify that the calculated hash of x (outCardCommit) 
    is equal to the inputted hash (playerCardCommit)
  */

  component mimc = MiMCSponge(1, 220, 1);
  mimc.ins[0] <== playerCard;
  mimc.k <== 0;

  outCardCommit <== mimc.outs[0];
  outCardCommit === playerCardCommit;

  /*
    Verify that player card is larger than threshold
    outValid = 1 if x less than threshold;
    outValid = 1 if x >= threshold
  */

  component greater = LessThan(11);
  greater.in[0] <== playerCard;
  greater.in[1] <== dealerCard;

  outValid <== greater.out; 
}

component main = Main();
\end{lstlisting}

\newpage
\printbibliography
\end{document}